\documentclass{article}
\usepackage{amsmath}
\usepackage{amssymb}
\usepackage{amsfonts}
\usepackage{amsthm}
\usepackage{bbm}
\usepackage{xcolor}
\usepackage{verbatim}
\usepackage{geometry}
\usepackage{booktabs}
\usepackage{multirow}
\usepackage{graphicx}
\usepackage{pdflscape}
\usepackage[hidelinks]{hyperref}
\usepackage{natbib}
\usepackage{authblk}
\usepackage{tabularx}

\newtheorem{example}{Example}
\newtheorem{definition}{Definition}

\newtheorem{theorem}{Theorem}[section]

\newtheorem{proposition}[theorem]{Proposition}
\newtheorem{corollary}[theorem]{Corollary}
\newtheorem{assumption}[theorem]{Assumption}

\title{Mixed Engression: Distributional Regression for Multivariate Mixed-Type Responses}

\author{Seyong Hwang}
\author{Kean Ming Tan}
\author{Ji Zhu}

\affil{Department of Statistics, University of Michigan, Ann Arbor}

\date{}

\begin{document}

\maketitle

\begin{abstract}
    Estimating the conditional distribution of multivariate responses with continuous and binary components requires capturing their dependence, respecting their different supports, and accommodating potentially heavy-tailed continuous outcomes. We propose mixed engression, which fits a continuous generator using energy losses induced by bounded discrepancies and then thresholds its soft binary outputs to produce mixed-response samples. Because the discrepancies are bounded, the population objective remains well defined without response-moment assumptions. However, we show that the full-response energy distance for the complete response vector can tend to zero before thresholding yet remain bounded away from zero afterward as response dimensions grow. A separate energy loss on the entire binary vector captures discrepancies obscured by the full-response comparison and controls the effect of thresholding. The generator is therefore fitted using an augmented energy loss that combines the full-response and binary components. Under compositional smoothness of the structural functions and a uniform threshold-margin condition, we derive nonasymptotic bounds on the final sampler’s full-response error for sparse ReLU estimators of models driven by Gaussian or Student-\(t\) structural noise, yielding polynomial convergence rates up to logarithmic factors. We also establish consistency when the covariate, response, and noise dimensions grow under suitable conditions. Simulations show improved accuracy relative to the comparison methods under the full-response energy distance, including in heavy-tailed settings. Moderate binary augmentation substantially improves binary-distribution accuracy while largely preserving full-response accuracy.
\end{abstract}

\section{Introduction}\label{sec:intro}


Distributional regression aims to estimate the conditional distribution of a response variable given covariates, rather than only a particular summary such as the conditional mean or quantiles \citep{kneib2023rage}.
The conditional distribution provides information about uncertainty, heterogeneity, and other features that cannot in general be recovered from a single conditional functional \citep{hothorn2014conditional}.
When the response is multivariate, estimating the joint conditional distribution also captures dependence among its coordinates.
A fitted conditional distribution provides a common basis for conditional sampling, prediction intervals, and distributional summaries, including conditional means when they exist.


Existing approaches to distributional regression estimate conditional distributions through collections of conditional quantiles \citep{koenker1978regression, JMLR:v7:meinshausen06a}, distribution functions \citep{hothorn2014conditional}, or densities \citep{dunson2007bayesian}. 
Generalized additive models for location, scale, and shape model covariate-dependent parameters of univariate distributions \citep{https://doi.org/10.1111/j.1467-9876.2005.00510.x,https://doi.org/10.1111/j.1467-9876.2011.01033.x}, while structured additive distributional regression extends this approach to multivariate responses \citep{10.1111/rssc.12090}. These approaches describe covariate effects, but their distributional flexibility remains tied to the chosen response family. 
Generative models offer an alternative by representing conditional distributions implicitly through a generator that transforms covariates and auxiliary noise into response samples, without requiring an explicit joint likelihood \citep{zhou2023deep}. Neural generators can be fitted using sample-based distributional discrepancies, including maximum mean discrepancy \citep{li2015generative,dziugaite2015training}, or proper scoring rules \citep{JMLR:v25:23-0038}. Engression \citep{shen2025engression} adopts this perspective by fitting a neural generator with an energy-score-based objective \citep{gneiting2007strictly, szekely2013energy}. The resulting framework is computationally simple, accommodates multivariate responses, and supports conditional sampling and the estimation of distributional summaries from a single fitted model.


Distributional regression for a multivariate response containing both continuous and binary coordinates requires learning dependence across variables with different supports. For example, jointly modeling blood pressure and diabetes status given age and body mass index requires more than estimating their separate conditional distributions. Latent-variable and copula models provide established approaches to representing such dependence \citep{sammel1997latent,murray2013bayesian,10.1093/biomet/93.3.537,https://doi.org/10.1002/sim.4087}.
The generative formulation of engression offers another approach to modeling the joint conditional distribution.
However, applying continuous-output neural generative models with the usual Euclidean energy loss in this setting raises two difficulties. First, the fitted conditional distribution must respect the binary support while representing dependence within and across response types; continuous-valued generator outputs do not by themselves enforce this support restriction. Second, continuous responses may be unbounded and heavy-tailed, whereas the usual expectation-based Euclidean energy loss and energy distance require finite first moments, which may not exist \citep{gneiting2007strictly}. 
These difficulties call for a distributional loss that remains well defined without response-moment assumptions and controls the accuracy of the mixed-response samples produced after fitting.

We propose mixed engression, which fits a continuous generator and thresholds its soft binary outputs to produce mixed-response samples. The fitted and target conditional distributions are compared using the full-response energy distance, a bounded energy distance for the complete response vector. However, as the response dimensions grow, a small full-response error of the soft generator need not ensure a small error for the thresholded sampler. An example shows how variation in the continuous output can mask missing binary variation: the full-response energy distance tends to zero before thresholding but remains bounded away from zero afterward. To address this failure, mixed engression fits the soft generator using an augmented energy loss that combines the full-response energy loss with a binary energy loss for the entire binary vector. The binary component detects the missing variation and controls the additional error introduced by thresholding. A threshold-transfer bound then controls the full-response error of the final sampler through the augmented error before thresholding.

Both loss components are induced by bounded Cauchy discrepancies, so their population risks remain well defined without response-moment assumptions. Recent finite-sample analyses of engression and related neural distributional regression establish their main guarantees under restrictive bounded-response assumptions, without developing theory specifically for binary responses \citep{huang2026theoretical,chen2026error,chai2026neuralgenerativedistributionalregression}. We develop statistical theory for mixed responses with binary components and unbounded continuous conditional support, including models without a finite conditional mean. Under compositional smoothness of the structural functions and a uniform threshold-margin condition that limits the concentration of latent binary scores near their thresholds, we derive nonasymptotic bounds for sparse ReLU estimators of models driven by shared Gaussian or Student-\(t\) structural noise.
With fixed dimensions and structural parameters, these bounds yield polynomial convergence rates, up to logarithmic factors, for the full-response energy distance of the final thresholded sampler. They also establish consistency when the covariate, response, and noise dimensions grow under suitable uniform regularity and growth conditions.

Simulations demonstrate the complementary effects of the two loss components. Fitting with the bounded full-response energy loss improves accuracy over standard engression under the full-response energy distance, with larger gains in the heavier-tailed settings considered. Binary augmentation substantially improves estimation of the binary distribution even when the full-response energy distance changes little. Moderate augmentation weights achieve these gains while largely preserving the full-response accuracy of the bounded-loss fit.

\subsection{Notation}\label{sec:notation}

For a random vector $Z$, let $P_Z$ denote its distribution, and write $Z \overset{d}{=}W$ for equality in distribution. We use $\Vert\cdot\Vert$ for the Euclidean norm and $\Vert v \Vert_\infty=\max_j|v_j|$ for the maximum norm of a vector $v$. For a measurable function $f:\mathcal{X} \to \mathbb{R}^d$, define
\[
\Vert f\Vert_\infty = \sup_{x \in \mathcal{X}} \Vert f(x) \Vert_\infty, \quad \Vert f \Vert_{\infty,2} = \sup_{x \in \mathcal{X}} \Vert f(x)\Vert.
\]
For a matrix $A=(A_{ij})$, let
\[
\Vert A \Vert_\infty = \max_{i,j} |A_{ij}|, \quad \Vert A \Vert_0 = \sum_{i,j} \mathbf{1}\{A_{ij} \neq0\},
\]
denote its elementwise maximum norm and the number of nonzero entries, respectively. For a generator $g(x,\xi)$ with auxiliary noise $\xi$ independent of $X$, let $P_{g \mid X=x}$ denote the distribution of $g(x,\xi)$, with $x$ fixed and randomness taken only over $\xi$. We write $P_{g \mid X}$ when the conditioning value is left implicit.
For nonnegative sequences $a_n$ and $b_n$, we write $a_n \lesssim b_n$ if $a_n \le Cb_n$ for all sufficiently large $n$, where $C>0$ is independent of $n$, and $a_n \asymp b_n$ if both $a_n \lesssim b_n$ and $b_n \lesssim a_n$.

\section{Preliminaries}\label{sec:prelim}

This section reviews energy distances and scores, introduces the standard engression formulation, and describes the sparse ReLU network class used in our method.

\subsection{Energy Statistics and Engression}\label{sec:Energy_stat}

Let $\mathcal{Y}$ be a measurable space, and let $\varphi: \mathcal{Y} \times \mathcal{Y} \to [0,\infty)$ be a measurable, symmetric function satisfying $\varphi(z,z)=0$.

\begin{definition}[Negative-type discrepancy]\label{def:neg_dis}
    The function $\varphi$ is of negative definite type if, for every integer $N \ge 2$, every $z_1,\ldots,z_N \in \mathcal{Y}$, and every $a_1,\ldots,a_N \in \mathbb{R}$ satisfying $\sum_{i=1}^N a_i=0$,
    \[
    \sum_{i=1}^N \sum_{j=1}^N a_ia_j\varphi(z_i,z_j) \le 0.
    \]
\end{definition}

\begin{definition}[Energy distance]\label{def:GED}
    Let $P$ and $Q$ be probability distributions on $\mathcal{Y}$. Let $Z,Z' \overset{\mathrm{i.i.d.}}{\sim} P$ and $ W,W' \overset{\mathrm{i.i.d.}}{\sim} Q$,
    with all random vectors mutually independent. The energy distance induced by $\varphi$ is
    \[
    \mathrm{ED}_\varphi(P,Q) = 2\mathbb{E}\varphi(Z,W) - \mathbb{E}\varphi(Z,Z') - \mathbb{E}\varphi(W,W'),
    \]
    whenever the expectations are finite.
\end{definition}

A negative-type discrepancy yields \(\mathrm{ED}_{\varphi}(P,Q)\ge 0\) whenever the defining expectations are finite. We say that \(\varphi\) is of strong negative type if, for every pair of probability distributions \(P\) and \(Q\) for which these expectations are finite,
\[
\mathrm{ED}_{\varphi}(P,Q)=0
\quad\Longleftrightarrow\quad
P=Q.
\]
In this case, the energy distance characterizes equality of distributions, and \(d_{\varphi}(P,Q)=\mathrm{ED}_{\varphi}(P,Q)^{1/2}\) is a metric on the corresponding class of probability distributions \citep{szekely2017energy,lyons2013distance,Sejdinovic_2013}. If \(\varphi\) is bounded, all defining expectations are finite for arbitrary probability distributions, so no moment condition is required.
The following example contrasts two strong negative-type discrepancies used in this paper: the unbounded Euclidean discrepancy and the bounded Cauchy discrepancy.

\begin{example}[Euclidean and Cauchy energy distances]
    On $\mathbb{R}^d$, the Euclidean discrepancy $\varphi_0(z,w)= \Vert z-w\Vert_2$ induces the Euclidean energy distance, denoted by $\mathrm{ED}(P,Q)$. This discrepancy is unbounded and of strong negative type, and its induced energy distance is well defined for distributions with finite first moments.

    In contrast, for an injective linear map $A : \mathbb{R}^d \to \mathbb{R}^r$, define the Cauchy discrepancy
    \begin{align}\label{eq:Cauchy_dis}
    \varphi_A(z,w) = \frac{\Vert A(z-w)\Vert_2^2}{1+\Vert A(z-w)\Vert_2^2}.
    \end{align}
    This discrepancy satisfies $0 \le \varphi_A(z,w) \le 1$ and is of strong negative type \citep{Sejdinovic_2013,JMLR:v13:gretton12a,JMLR:v11:sriperumbudur10a}. Consequently, the induced Cauchy energy distance is well defined and characterizes equality of arbitrary probability distributions on $\mathbb{R}^d$ without moment assumptions.
\end{example}

\begin{definition}[Energy score]\label{def:GES}
    Let $P$ be a probability distribution on $\mathcal{Y}$, and let $z \in \mathcal{Y}$. If $Z,Z'\overset{\mathrm{i.i.d.}}{\sim} P$, the energy score induced by $\varphi$ is
    \[
    \mathrm{ES}_\varphi(P,z) = \frac{1}{2}\mathbb{E}[\varphi(Z,Z')] - \mathbb{E}[\varphi(Z,z)],
    \]
    provided both expectations are finite.
\end{definition}
Larger values indicate better distributional predictions, so estimation proceeds by minimizing the negative score.

If $Y,Y' \overset{\mathrm{i.i.d.}}{\sim} Q$, then
\begin{align}\label{eq:equiv_ed_es}
    \mathbb{E}_{Y \sim Q}[-\mathrm{ES}_\varphi(P,Y)] = \frac{1}{2}\mathrm{ED}_\varphi(P,Q)+\frac{1}{2}\mathbb{E}[\varphi(Y,Y')].
\end{align}
The second term in \eqref{eq:equiv_ed_es} does not depend on $P$. Hence, minimizing the expected negative energy score over $P$ is equivalent to minimizing $\mathrm{ED}_\varphi(P,Q)$. When $\varphi$ is of strong negative type, the score is strictly proper on the same class of distributions.


\citet{shen2025engression} introduce engression as a generative distributional regression method based on the Euclidean energy score. This corresponds to the Euclidean discrepancy $\varphi_0(z,w)=\Vert z-w \Vert_2$. For distributions $P$ and $Q$ on $\mathbb{R}^d$ with finite first moments,
\[
\mathrm{ED}(P,Q) = 2\mathbb{E}\Vert Z-W\Vert_2 - \mathbb{E}\Vert Z-Z'\Vert_2-\mathbb{E}\Vert W-W'\Vert_2,
\]
where $Z,Z' \overset{\mathrm{i.i.d.}}{\sim} P$ and $W,W' \overset{\mathrm{i.i.d.}}{\sim} Q$ are mutually independent.

Let $\xi \sim P_\xi$ be independent of $(X,Y)$, and let $g:\mathcal{X} \times \Xi \to \mathbb{R}^d$ belong to a generator class $\mathcal{G}$. Denote by $P_{g \mid X=x}$ the distribution of $g(x,\xi)$. Taking an independent copy $\xi'$ of $\xi$, the population engression objective is
\[
\arg\min_{g \in \mathcal{G}} \mathbb{E}\left[ \Vert Y-g(X,\xi)\Vert_2 - \frac{1}{2}\Vert g(X,\xi)-g(X,\xi')\Vert_2 \right] = \arg\min_{g \in \mathcal{G}} \mathbb{E}_X[\mathrm{ED}(P_{g \mid X},P_{Y \mid X})].
\]
Under correct specification, if the minimum is attained, every population minimizer satisfies $P_{g \mid X=x}=P_{Y \mid X=x}$ for $P_X$-almost every $x$. Its empirical version replaces the expectation over $(X,Y)$ with a sample average and the expectations over the auxiliary noise with Monte Carlo averages.
In the remainder of this paper, we refer to this original formulation as standard engression.

Section~\ref{sec:methodology} introduces mixed engression, which retains this scoring-rule formulation but replaces the Euclidean discrepancy with a bounded discrepancy tailored to mixed responses.

\subsection{Sparse ReLU Generators}\label{sec:ReLU}

We parametrize candidate generators using sparse rectified linear unit (ReLU) networks.
Let $L \ge 1$ denote the depth, and let $\boldsymbol{p}=(p_0,\ldots,p_{L+1})^\top$ be a width vector. A ReLU network with architecture $(L,\boldsymbol{p})$ is a map $g:\mathbb{R}^{p_0} \to \mathbb{R}^{p_{L+1}}$ of the form
\begin{align*}
    g(x) = W_L\sigma_{v_L}W_{L-1}\sigma_{v_{L-1}} \cdots W_1\sigma_{v_1}W_0x,
\end{align*}
where $W_\ell \in \mathbb{R}^{p_{\ell+1} \times p_{\ell}}$ for $\ell=0,\ldots,L$, $v_\ell \in \mathbb{R}^{p_\ell}$ for $\ell=1,\ldots,L$, and $\sigma_v(u)=\max\{u+v,0\}$ is applied coordinatewise.

\begin{definition}[Sparse ReLU network class]\label{def:ReLU}
    For a sparsity bound $s$, let $\mathcal{G}(L,\boldsymbol{p},s)$ be the class of maps admitting the above representation with
    \[
    \sum_{\ell=0}^L \Vert W_\ell\Vert_0 + \sum_{\ell=1}^L\Vert v_\ell\Vert_0 \le s, \quad \max_{0 \le \ell \le L} \Vert W_\ell\Vert_\infty \vee \max_{1 \le \ell \le L} \Vert v_\ell\Vert_\infty \le 1.
    \]
\end{definition}

For mixed-response generation, the network input is $(x,\xi)$, so $p_0=p+k$ and $p_{L+1}=d$. 
The estimator in Section~\ref{sec:sample_gen_engression} is obtained by minimizing the empirical energy loss over such a sparse ReLU network class. The approximation properties and covering-number bounds for this class, established in \cite{Schmidt_Hieber_2020}, form the basis for the consistency analysis in the following sections.

\section{Mixed Engression}\label{sec:methodology}

This section introduces mixed engression for multivariate mixed-type responses. We first specify the response model and distinguish the soft generator used for fitting from the thresholded sampler. We then define the bounded discrepancies underlying the augmented energy loss, establish population guarantees connecting the fitting objective to the thresholded sampling distribution, and formulate the finite-sample estimator.

\subsection{Mixed-Response Model}\label{sec:mixed_model}

Let $X \in \mathcal{X} \subset \mathbb{R}^p$ be the $p$-dimensional covariate vector, and let
\[
Y = \begin{pmatrix} Y^c \\ Y^b \end{pmatrix} \in \mathbb{R}^{d_c} \times \{0,1\}^{d_b},\quad d=d_c+d_b
\]
be a $d$-dimensional response vector,
where $Y^c$ and $Y^b$ are the continuous and binary components of $Y$, respectively.
The goal is to estimate the full conditional distribution $P_{Y \mid X}$. 
When evaluating the energy distances, we regard $Y$ as a random vector in $\mathbb{R}^d$, with the binary coordinates supported on $\{0,1\}$.

We impose the following structural assumption on the true conditional distribution.

\begin{assumption}[Structural mixed-response representation]\label{assu:mixed_model}
    There exist structural noise variables $\epsilon =(\epsilon_1,\ldots,\epsilon_k)^{\top} $, independent of $X$, with $\epsilon_j \overset{\mathrm{i.i.d.}}{\sim} \nu_\epsilon$, and jointly continuous functions 
    \[
    f^c : \mathcal{X}\times\mathbb{R}^k \to \mathbb{R}^{d_c}, \quad f^b:\mathcal{X}\times\mathbb{R}^{k} \to \mathbb{R}^{d_b},
    \]
    such that
    \begin{align}\label{eq:mixed_representation}
        (Y^c,Y^b) = \left(f^c(X,\epsilon),\mathbf{1}\{f^b(X,\epsilon) \ge 0\}\right),
    \end{align}
    where $\nu_\epsilon$ has a strictly positive density on $\mathbb{R}$, and the indicator is applied coordinatewise.
\end{assumption}

The shared structural noise $\epsilon$ allows the continuous and binary components to be conditionally dependent given $X$. Specifically, the representation in~\eqref{eq:mixed_representation} accommodates dependence both within and across the continuous and binary response blocks.

Although Assumption~\ref{assu:mixed_model} provides a representation of the target distribution $P_{Y\mid X}$, it raises two challenges for estimation.  
First, the above representation is generally non-unique, i.e., different structural noise distributions and transformation functions may induce the same conditional distribution.  In other words, the functions $f^c$ and $f^b$ and the distribution of $\epsilon$ are neither identifiable nor targets of estimation. Second, the binary component is generated by the discontinuous map $\mathbf{1}\{f^b(X,\epsilon) \ge 0 \}$. Directly incorporating this thresholding operation into the estimation criterion would lead to a discontinuous optimization problem.

To address these challenges, Section~\ref{sec:soft_thres_gen} introduces a standardized auxiliary noise variable and a continuous soft generator. The soft generator is used for optimization, while a thresholded version is used to produce the final mixed-type distribution.

\subsection{Soft Generator and Thresholded Sampler}\label{sec:soft_thres_gen}

We begin by expressing the structural noise in Assumption~\ref{assu:mixed_model} using uniform auxiliary noise. Let $\xi \sim \mathrm{Unif}([0,1]^k)$ be independent of $X$, and let $F_{\nu_\epsilon}$ denote the cumulative distribution function of $\nu_\epsilon$. Define
\[
\Psi(u) = \left( F_{\nu_\epsilon}^{-1}(u_1),\ldots,F_{\nu_\epsilon}^{-1}(u_k) \right)^\top, \quad u \in (0,1)^k,
\]
and set $\Psi(u)=0$ on the boundary of $[0,1]^k$, which has probability zero under the distribution of $\xi$. Then $\Psi(\xi)\overset{d}{=} \epsilon$. Replacing $\epsilon$ by $\Psi(\xi)$ in \eqref{eq:mixed_representation} therefore preserves the conditional distribution $P_{Y \mid X}$, motivating the use of the common input $(X,\xi)$ for candidate generators.

A candidate soft generator is a map
\[
g=(g^c,g^b),\quad g^c:\mathcal{X}\times[0,1]^k \to \mathbb{R}^{d_c},\quad g^b:\mathcal{X}\times [0,1]^k \to [0,1]^{d_b},
\]
where $g^c$ and $g^b$ correspond to the continuous and binary response blocks, respectively.
We refer to $g$ as a soft generator because the binary block $g^b$ takes values in the relaxed cube $[0,1]^{d_b}$ and is used directly in the optimization criterion.
Let $\mathcal{G}$ denote a class of candidate soft generators of the above form. Unless otherwise stated, elements of $\mathcal{G}$ are measurable maps from $\mathcal{X}\times[0,1]^k$ into $\mathbb{R}^{d_c} \times [0,1]^{d_b}$.
For any $x \in \mathcal{X}$, let $P_{g \mid X=x}$ be the conditional distribution of the soft output
\[
g(x,\xi) = \left(g^c(x,\xi),g^b(x,\xi)\right),
\]
where the randomness arises only from $\xi$. 
The output $g^b(x,\xi)$ takes values in $[0,1]^{d_b}$ and serves as a continuous surrogate for the binary response. To recover a mixed-type output, define the coordinatewise thresholding operation $T_{1/2}(u)_j=\mathbf{1}\{u_j \ge 1/2\}$, and let $\mathcal{T}$ denote the thresholding operator on $g$ (acting on the binary block):
\[
(\mathcal{T}g)(x,\xi) = \left(g^c(x,\xi),T_{1/2}(g^b(x,\xi))\right).
\]
The corresponding conditional distribution is denoted by $P_{\mathcal{T}g \mid X=x}$.

Recall that the goal is not to recover the true generators $f^c$ or $f^b$. Mixed engression instead searches for a soft generator $g$ whose induced conditional distribution $P_{g \mid X=x}$ approximates the true conditional distribution $P_{Y \mid X=x}$. Optimization is carried out using the soft output $g(X,\xi)$, so $P_{g \mid X=x}$ is the soft generator distribution used in the population and empirical objectives. After a generator has been fitted, the thresholding operator $\mathcal{T}$ converts the binary block of the soft output into samples in $\{0,1\}^{d_b}$. Thus $P_{\mathcal{T}g \mid X}$ is the final mixed-type sampling distribution.

The following example illustrates that the structural and fitted generators can have different functional forms and noise distributions while inducing the same conditional distribution after thresholding.

\begin{example}[Different generators with the same thresholded distribution]\label{ex:ex1}
Consider a scalar binary response, and let
\[
f^b(x,\epsilon) = x+\epsilon, \quad \epsilon \sim N(0,1).
\]
Then
\[
(Y^b \mid X=x) = \mathbf{1}\{x+\epsilon \ge 0\} \sim \mathrm{Bernoulli}\left(\Phi_0(x)\right),
\]
where $\Phi_0$ is the standard normal distribution function. The same family of thresholded conditional distributions can be generated using a different auxiliary noise variable. Let $\xi \sim \mathrm{Unif}(0,1)$ and define
\[
g^b(x,\xi) = \frac{\xi + \Phi_0(x)}{2} \in [0,1].
\]
Then
\[
T_{1/2}\{g^b(x,\xi)\} =\mathbf{1}\{\xi \ge 1-\Phi_0(x)\}\sim \mathrm{Bernoulli}\left(\Phi_0(x)\right).
\]
Hence $P_{\mathcal{T}g \mid X=x}=P_{Y \mid X=x}$ for every $x$, although $f^b$ and $g^b$ are different maps and $\epsilon$ and $\xi$ have different distributions. Notice also that the unthresholded variable $g^b(x,\xi)$ is not Bernoulli; only the thresholded distribution $P_{\mathcal{T}g \mid X=x}$ agrees with the target binary conditional distribution.
\end{example}

\subsection{Mixed-Response Discrepancies and Energy Distances}\label{sec:mixed_ED}

Section~\ref{sec:Energy_stat} introduced energy distances induced by negative-type discrepancies. We now 
introduce energy distances induced by discrepancies tailored to mixed-type responses. Unlike a direct application of the Euclidean energy distance to the full response vector, the proposed construction explicitly accounts for the different roles of the continuous and binary blocks. Our construction therefore combines two complementary components, $\varphi_F$ and $\varphi_B$, defined in Definition~\ref{def:mixed_dis}.
Let $\rho(t) = t/(1+t)$ for $t \ge 0$.

\begin{definition}[Mixed-response discrepancies]\label{def:mixed_dis}
Let $d_c,d_b \ge 1$, and let $y=(y^c,y^b)$ and $z=(z^c,z^b)$ belong to $\mathbb{R}^{d_c} \times [0,1]^{d_b}$.
    Define the full-response and binary discrepancies by
    \[
    \varphi_F(y,z) = \rho\left( \frac{\Vert y^c-z^c\Vert_2^2}{d_c} + \frac{\Vert y^b-z^b\Vert_2^2}{d_b} \right), \quad \varphi_B(y^b,z^b) = \rho\left( \frac{\Vert y^b-z^b\Vert_2^2}{d_b} \right).
    \]
    For a fixed weight $\omega \in [0,1)$, define the augmented discrepancy by
    \begin{align}\label{eq:aug_dis}
            \Phi(y,z) = (1-\omega)\varphi_F(y,z) + \omega\varphi_B(y^b,z^b).
        \end{align}
    The induced energy distances are denoted by $\mathrm{ED}_F$, $\mathrm{ED}_B$, and $\mathrm{ED}_\Phi$, respectively, and the corresponding energy scores by $\mathrm{ES}_F$, $\mathrm{ES}_B$, and $\mathrm{ES}_\Phi$.
\end{definition}

The full-response discrepancy $\varphi_F$ is a block-normalized Cauchy discrepancy of the form \eqref{eq:Cauchy_dis} on the complete response vector, after the injective linear rescaling $y \mapsto (y^c/\sqrt{d_c}, y^b/{\sqrt{d_b}})$. The separate normalization by $d_c$ and $d_b$ keeps the two response blocks on comparable average-coordinate scales and prevents either block from dominating the geometry solely because it contains more coordinates. 
The full-response energy distance assesses the full response distribution, including dependence within and across the blocks, whereas the binary energy distance assesses the binary distribution.
Finally, the augmented discrepancy $\Phi$ is defined as a weighted sum of the two discrepancies. The discrepancies \(\varphi_F\) and \(\Phi\) are bounded and of strong negative type on \(\mathbb R^{d_c}\times[0,1]^{d_b}\), while \(\varphi_B\) is bounded and of strong negative type on \([0,1]^{d_b}\).

\begin{proposition}[Decomposition and identification]\label{prop:identification}
    For probability measures $P$ and $Q$ on $\mathbb{R}^{d_c} \times [0,1]^{d_b}$, let $P^b$ and $Q^b$ denote the marginal distributions of their binary components. For $z = (z^c,z^b) \in \mathbb{R}^{d_c}\times[0,1]^{d_b}$ in the same space,
    \begin{align*}
        \mathrm{ED}_\Phi (P,Q)  &= (1-\omega)\mathrm{ED}_F(P,Q) + \omega\mathrm{ED}_B(P^b,Q^b), \\
        \mathrm{ES}_\Phi (P,z) &=(1-\omega)\mathrm{ES}_F(P,z) + \omega\mathrm{ES}_B(P^b,z^b).
    \end{align*}
    Moreover, for every fixed finite $d_c,d_b$ and $\omega \in [0,1)$,
    \[
    \mathrm{ED}_F(P,Q) = 0 \Longleftrightarrow \mathrm{ED}_\Phi(P,Q)=0 \Longleftrightarrow P=Q.
    \]
\end{proposition}

The decompositions follow from the linearity of the energy distance and score in the discrepancy. Identification follows from the strong negative type of the full-response discrepancy and its positive weight in the augmented discrepancy.
The full-response energy distance alone is sufficient for population identification at any fixed response dimension. However, the following example shows that, as the response dimension grows, the full-response energy distance between the soft and target distributions can vanish while the corresponding distance after thresholding remains bounded away from zero. The binary energy distance detects the missing binary variation in this example, motivating its inclusion to provide quantitative control of the thresholded sampler.

\begin{example}[Failure of full-response energy distance control after thresholding]\label{ex:ex2}
Let $d_c=d_b=r$, fix a Bernoulli probability $p \in (0,1/2)$, and let $s=p(1-p)$. Let $B_r$ have independent $\mathrm{Bernoulli}(p)$ coordinates, and define the target distribution
\[
P_r=\mathcal{L}(0_r,B_r),
\]
where $\mathcal{L}$ denotes the distribution of a random vector. 
In this example, $\mathrm{ED}_{F,r}$ and $\mathrm{ED}_{B,r}$ denote the energy distances in Definition~\ref{def:mixed_dis} with $d_c=d_b=r$.
For $\xi_r \sim \mathrm{Unif}([0,1]^{r})$, define
\[
U_r = \sqrt{12s}\left( \xi_r-\frac{1}{2}\mathbf{1}_r \right), \quad g_r(\xi_r) = (U_r,p\mathbf{1}_r),\quad Q_r =\mathcal{L}\{g_r(\xi_r)\}.
\]
The binary block of the soft generator $g_r$ is constant at $p$ and therefore has no binary variation. Nevertheless, as $r \to \infty$,
\[
\mathrm{ED}_{F,r}(P_r,Q_r) \to 0.
\]
Since $p<1/2$, thresholding sends every binary coordinate to zero. Thus, for $\mathcal{T}Q_r = \mathcal{L}(U_r,0_r)$, 
\[
\mathrm{ED}_{F,r}(P_r,\mathcal{T}Q_r) \to 2\{\rho(s+p)-\rho(2s)\}>0.
\]
In contrast, the binary energy distance satisfies
\[
\mathrm{ED}_{B,r}(P_r^b,Q_r^b) \to 2\rho(s)-\rho(2s) = \frac{2s^2}{(1+s)(1+2s)} > 0.
\]
Thus, the full-response energy distance can conceal the missing binary variation through the continuous block, whereas the binary energy distance detects this failure at a nonvanishing scale.
\end{example}

\subsection{Population-Level Estimation}\label{sec:population}

\subsubsection{Population Criterion}\label{sec:pop_criterion}

For the population analysis, we consider the following class of candidate soft generators:
\[
\mathcal{G}_{\mathrm{pop}} = \{ g: \mathcal{X}\times[0,1]^k \to \mathbb{R}^{d_c} \times [0,1]^{d_b}: g \text{ is measurable and continuous on } \mathcal{X} \times(0,1)^k\}.
\]
The values on the boundary of $[0,1]^k$ are immaterial because $\xi \sim \mathrm{Unif}([0,1]^k)$ assigns probability zero to the boundary. This formulation retains the compact-noise notation used by the estimator while allowing generator maps that may become unbounded as $\xi$ approaches the boundary.

For $g \in \mathcal{G}_{\mathrm{pop}}$, define the population risk as the expected negative augmented energy score between the conditional distribution induced by the soft generator, $P_{g \mid X}$, and the response $Y$:
\[
\mathcal{R}_\Phi(g) = \mathbb{E}_{X,Y}\left[-\mathrm{ES}_\Phi(P_{g \mid X},Y)\right] = \mathbb{E}_{X,Y,\xi} \left[\Phi\left(g(X,\xi),Y\right)\right] - \frac{1}{2}\mathbb{E}_{X,\xi,\xi'}\left[\Phi\left(g(X,\xi),g(X,\xi')\right)\right],
\]
where $\xi$ and $\xi'$ are independent copies of $\mathrm{Unif}([0,1]^k)$ and are independent of $(X,Y)$.

We define the augmented error of $g$ as the expected augmented energy distance between the soft generator and target conditional distributions:
\[
\mathcal{E}_\Phi(g) = \mathbb{E}_X[\mathrm{ED}_\Phi(P_{g \mid X},P_{Y \mid X})].
\]
By \eqref{eq:equiv_ed_es}, the population risk satisfies
\begin{align}\label{eq:ED_and_risk}
    \mathcal{R}_\Phi(g) = \frac{1}{2}\mathbb{E}_X[\mathrm{ED}_\Phi(P_{g\mid X},P_{Y \mid X})] + \frac{1}{2}\mathbb{E}_X\mathbb{E}[\Phi(Y,Y') \mid X],
\end{align}
where $Y'$ is a conditionally independent copy of $Y$ given $X$.
Since $\mathbb{E}_X\mathbb{E}[\Phi(Y,Y') \mid X]$ does not depend on $g$, minimizing the population risk $\mathcal{R}_\Phi(g)$ is equivalent to minimizing the augmented error $\mathcal{E}_\Phi(g)$.
By Proposition~\ref{prop:identification}, the augmented error is non-negative and equals zero if and only if $P_{g \mid X=x}=P_{Y \mid X=x}$ for $P_X$-almost every $x$.

However, because minimizers of $\mathcal{R}_\Phi(g)$ (or $\mathcal{E}_\Phi(g)$) may fail to exist in mixed-response settings with continuous soft-generator classes, the primary population objective is the infimum criterion:
\[
\mathcal{E}_\Phi^* = \inf_{g \in \mathcal{G}_{\mathrm{pop}}} \mathcal{E}_\Phi(g).
\]

\subsubsection{Population Guarantee and Threshold-Transfer}\label{sec:pop_guarantee}

The population criterion in Section~\ref{sec:pop_criterion} is defined over the soft conditional distributions $P_{g \mid X}$. In the mixed-response setting, the infimum augmented error is zero, i.e., $\mathcal{E}^*_\Phi=0$, but a continuous soft generator generally cannot attain this infimum exactly when a binary coordinate is conditionally non-degenerate.

\begin{theorem}[Zero infimum and non-attainment]\label{thm:zero_inf}
    Under Assumption~\ref{assu:mixed_model}, $\mathcal{E}^*_\Phi=0$.
    If, for at least one binary coordinate $j$,
    \[
    P_X\{x:0<\mathrm{Pr}(Y^b_j=1 \mid X=x)<1\} > 0,
    \]
    then the zero infimum above is not attained by any $g \in \mathcal{G}_{\mathrm{pop}}$.
\end{theorem}

The non-attainment result reflects a mismatch between the soft optimization space and the mixed support of the target distribution. When a binary coordinate is conditionally non-degenerate, the target distribution assigns positive mass to both binary values. A continuous soft generator can approximate such a distribution arbitrarily well by making its binary output increasingly close to 0 or 1, but it cannot exactly produce a non-degenerate $\{0,1\}$-valued distribution while remaining continuous in the auxiliary noise. Thus, in the mixed-response setting, the population criterion should be interpreted as an infimum problem over soft conditional distributions rather than as an optimization problem with an exact minimizer.

To assess the effect of thresholding, define the soft full-response, binary, and thresholded full-response errors by
\begin{align*}
\mathcal{E}_F(g) &= \mathbb{E}_X\mathrm{ED}_F(P_{g \mid X}, P_{Y \mid X}), \\
\mathcal{E}_B(g) &= \mathbb{E}_X\mathrm{ED}_B(P_{g^b \mid X},P_{Y^b \mid X}), \\
\mathcal{E}_F(\mathcal{T}g) &= \mathbb{E}_X\mathrm{ED}_F(P_{\mathcal{T}g \mid X}, P_{Y \mid X}).
\end{align*}
Proposition~\ref{prop:identification} implies
\[
\mathcal{E}_\Phi(g) = (1-\omega)\mathcal{E}_F(g) + \omega\mathcal{E}_B(g).
\]
Thus, the augmented error $\mathcal{E}_\Phi(g)$ controls the two soft-distribution errors $\mathcal{E}_F(g)$ and $\mathcal{E}_B(g)$, whereas the final sampler is evaluated through $\mathcal{E}_F(\mathcal{T}g)$. The next result relates this thresholded full-response error to the errors controlled during fitting.

\begin{theorem}[Binary threshold transfer]\label{thm:thres_transfer}
    Fix $\omega \in (0,1)$. For any $g \in \mathcal{G}_\mathrm{pop}$,
    \begin{align}\label{eq:thres_transfer}
        \mathcal{E}_F(\mathcal{T}g) \le 2\mathcal{E}_F(g) + \frac{25}{4}\{\mathcal{E}_B(g)\}^{1/2} \le \frac{2}{1-\omega}\mathcal{E}_\Phi(g) + \frac{25}{4\sqrt{\omega}} \{\mathcal{E}_\Phi(g)\}^{1/2}.
    \end{align}
\end{theorem}

The two terms in \eqref{eq:thres_transfer} have complementary interpretations. The soft full-response error $\mathcal{E}_F(g)$ controls whether the fitted soft distribution has learned the complete dependence structure of the target. The binary error $\mathcal{E}_B(g)$ controls concentration near the binary support and, consequently, the additional error introduced by thresholding. 

\begin{corollary}[Transfer from the objective to the full-response target]\label{cor:pop_thres_transfer}
    For any sequence $g_\ell \in \mathcal{G}_{\mathrm{pop}}$, 
    \[
    \mathcal{E}_\Phi(g_\ell) \to 0  \quad \Longrightarrow \quad \mathcal{E}_F(g_\ell) \to 0 \quad \text{and} \quad \mathcal{E}_F(\mathcal{T}g_\ell) \to 0.
    \]
\end{corollary}

Together with Theorem~\ref{thm:zero_inf}, this result shows that sequences approaching the population infimum are consistent for the final thresholded sampling distribution without requiring the existence of an exact population minimizer.

\subsection{Finite-Sample Estimator}\label{sec:sample_gen_engression}

Let $(X_i,Y_i)_{i=1}^n$ be i.i.d. observations from $P_{X,Y}$. 
We use the sparse ReLU class $\mathcal{G}(L,\boldsymbol{p},s)$ from Section~\ref{sec:ReLU}, with input dimension $p+k$ and output dimension $d=d_c+d_b$, and define
\[
\mathcal{G}_\mathrm{fin} = \{(h^c,\Pi_{[0,1]^{d_b}}h^b): h=(h^c,h^b) \in \mathcal{G}(L,\boldsymbol{p},s)\} \subseteq \mathcal{G}_\mathrm{pop},
\]
where the maps are restricted to $\mathcal{X} \times[0,1]^k$. The projection is applied coordinatewise using
\[
\Pi_{[0,1]}(t) = [t]_+ - [t-1]_+, \quad [t]_+ = \max\{t,0\}.
\]
This projection has an exact ReLU representation and enforces the relaxed binary range. The architecture and sparsity are specified in the finite-sample analysis in Section~\ref{sec:statistical_theory}.

For each observation $i$, draw 
\[
\xi_{i1},\ldots,\xi_{im} \overset{\mathrm{i.i.d.}}{\sim} \mathrm{Unif}([0,1]^k), \quad m \ge 2,
\]
independently across observations and independently of the data. 
For $g \in \mathcal{G}_\mathrm{fin}$, the Monte Carlo approximation to the negative augmented energy score at observation $i$ is
\[
-\widehat{\mathrm{ES}}_{\Phi}\left(P_{g \mid X=X_i},Y_i\right) = \frac{1}{m}\sum_{j=1}^m \Phi\left(g(X_i,\xi_{ij}),Y_i\right) - \frac{1}{2m(m-1)} \sum_{j \neq j'} \Phi\left(g(X_i,\xi_{ij}),g(X_i,\xi_{ij'})\right).
\]
The empirical Monte Carlo risk is defined as
\[
\hat{\mathcal{R}}_{n,m}(g) = \frac{1}{n}\sum_{i=1}^n [-\widehat{\mathrm{ES}}_\Phi(P_{g \mid X=X_i},Y_i)],
\]
and the finite-sample mixed engression estimator is its minimizer:
\begin{align}\label{eq:finite_gen_engression}
    \hat{g}_{n,m} \in \arg\min_{g \in \mathcal{G}_\mathrm{fin}} \hat{\mathcal{R}}_{n,m}(g).
\end{align}

The empirical risk $\hat{\mathcal{R}}_{n,m}$ approximates the population risk $\mathcal{R}_\Phi(g)$ in two respects: the expectation over $(X,Y)$ is replaced by the sample average, and the expectations over the auxiliary noise are replaced by Monte Carlo averages. As in the population criterion, optimization is performed using the soft generator outputs $\{g(X_i,\xi_{ij})\}$. The thresholding operator $\mathcal{T}$ is not applied within the empirical risk. After fitting, the estimated mixed-response conditional distribution is
\[
\hat{P}_{Y \mid X=x} = P_{\mathcal{T}\hat{g}_{n,m} \mid X=x}.
\]

Section~\ref{sec:statistical_theory} establishes bounds for its thresholded full-response error $\mathcal{E}_{F}(\mathcal{T}\hat{g}_{n,m})$ by first controlling the augmented error $\mathcal{E}_\Phi(\hat{g}_{n,m})$.

\section{Finite-Sample Theory for Mixed Engression}\label{sec:statistical_theory}

In this section, we establish finite-sample guarantees for the estimator $\hat{g}_{n,m}$ defined in Section~\ref{sec:sample_gen_engression}. 
We first introduce function classes and structural assumptions that accommodate both unbounded continuous responses and thresholded binary responses. 
We then derive a nonasymptotic bound for $\mathcal{E}_\Phi(\hat{g}_{n,m})$, the augmented error of the estimator, and apply the threshold-transfer inequality to bound the full-response error of the final mixed-response sampler $\mathcal{T}\hat{g}_{n,m}$.
We consider Gaussian and Student-$t$ structural-noise representations and derive high-dimensional consistency when the covariate, response, and noise dimensions grow.

Throughout this section, we take the covariate domain to be $\mathcal{X}=[0,1]^p$, a standard setting in nonparametric regression theory \citep{Schmidt_Hieber_2020}. The unit-cube normalization is for convenience: any fixed nondegenerate rectangular covariate domain can be reduced to this setting by coordinatewise affine rescaling, with corresponding changes to the regularity constants. The bounded-domain assumption concerns only the covariates and does not require the continuous responses to have bounded support.
We consider the estimator in \eqref{eq:finite_gen_engression} using the clipped sparse ReLU class $\mathcal{G}_\mathrm{fin}$ introduced in Section~\ref{sec:sample_gen_engression}, with a fixed augmentation weight $\omega \in (0,1)$.

\subsection{Compositional Smoothness for Mixed Responses}\label{sec:comp_smooth}

We describe the regularity of the structural continuous generator and the latent binary score using compositional Hölder classes. Following \cite{Schmidt_Hieber_2020}, each coordinate function in a composition is allowed to depend on only a limited number of inputs. We first introduce bounded Hölder balls and their compositional counterparts, which provide the framework for the ReLU approximation results used in our analysis. We then relax the uniform bound on function values and allow unbounded component domains to accommodate unbounded continuous responses. The smoothness assumptions for the binary component concern the latent score before thresholding.

Throughout this subsection, $\lfloor\beta\rfloor$ denotes the largest integer strictly smaller than $\beta$. For a multi-index $\alpha$, write $|\alpha| = \sum_j \alpha_j$.

\begin{definition}[Bounded Hölder ball]\label{def:bdd_Holder_ball}
    For $D \subseteq \mathbb{R}^t$, $\beta>0$, and $K>0$, the class $\mathcal{C}_t^\beta(D,K)$ consists of functions $f:D \rightarrow \mathbb{R}$ whose partial derivatives through order $\lfloor\beta\rfloor$ exist and satisfy 
    \begin{align}\label{eq:holder_ball}
        \sum_{\boldsymbol\alpha:|\boldsymbol\alpha| < \beta} \Vert \partial^{\boldsymbol\alpha}f\Vert_\infty + \sum_{\boldsymbol\alpha:|\boldsymbol\alpha|  =\lfloor\beta\rfloor} \sup_{x,y \in D, x\neq y}\frac{|\partial^{\boldsymbol\alpha}f(x) - \partial^{\boldsymbol\alpha}f(y)|}{\Vert x-y\Vert^{\beta - \lfloor\beta\rfloor}_\infty} \le K.
    \end{align}
\end{definition}

Here, \(t\) is the input dimension, \(\beta\) is the smoothness index, and \(K\) bounds the Hölder norm, including the supremum norm of the function.
To define the compositional classes, let $q \ge 0$ and write
\[
\boldsymbol{d} = (d_0,\ldots,d_{q+1})^\top, \quad \boldsymbol{t} = (t_0,\ldots,t_q)^\top, \quad \boldsymbol{\beta} = (\beta_0,\ldots,\beta_q)^\top, \quad \boldsymbol{K}=(K_0,\ldots,K_q)^\top,
\]
where $1 \le t_i \le d_i$, $\beta_i>0$, and $K_i>0$. 
Here, \(q+1\) is the number of component maps in the composition, and \(\boldsymbol{d} \) specifies the input, intermediate, and output dimensions. At stage \(i\), the component map takes \(d_i\) inputs and produces \(d_{i+1}\) outputs. Each output coordinate depends on at most \(t_i\) of these inputs and has Hölder smoothness \(\beta_i\). In the bounded compositional class defined below, \(K_i\) bounds the Hölder norm of each coordinate function and the absolute values of the input-domain endpoints.
For a coordinate set $S$, let $z_S$ denote the corresponding subvector and $\pi_S(D)$ the coordinate projection of a domain $D$.

\begin{definition}[Bounded Hölder compositional class]\label{def:bdd_Holder_comp}
    Let $I_i=[a_i,b_i]^{d_i}$ for $i=0,\ldots,q+1$. The class $\mathcal{F}(q,\boldsymbol{d},\boldsymbol{t},\beta,\boldsymbol{K})$ consists of maps admitting a representation
    \[
    f = f_q \circ\cdots\circ f_0, \quad f_i : I_{i} \to I_{i+1}, \quad i=0,\ldots,q,
    \]
    with $\max\{|a_i|,|b_i|\} \le K_i$ for $i=0,\ldots,q$. For every layer $i$ and coordinate $j=1,\ldots,d_{i+1}$, there exist a fixed set $S_{i,j} \subseteq \{1,\ldots,d_i\}$, with $1 \le |S_{i,j}| \le t_i$, and a function $\tilde{f}_{i,j}$ such that
    \[
    f_{i,j}(z) = \tilde{f}_{i,j}(z_{S_{i,j}}),\quad \tilde{f}_{i,j} \in \mathcal{C}_{|S_{i,j}|}^{\beta_i}\left(\pi_{S_{i,j}}(I_i),K_i\right).
    \]
\end{definition}

The selected coordinate sets may differ across layers and components but do not depend on the input value.
The term $\alpha=0$ in Definition~\ref{def:bdd_Holder_ball} imposes $\Vert f\Vert_\infty \le K$. Consequently, a generator in the bounded compositional class cannot produce an unbounded continuous response, even when the structural noise has unbounded support. We therefore remove only the uniform bound on the function itself while retaining the derivative and Hölder controls.

\begin{definition}[Unbounded Hölder ball]\label{def:unbdd_Holder_ball}
    For $D \subseteq \mathbb{R}^t$, $\beta>0$, and $K>0$, the class $\mathcal{D}_t^\beta(D,K)$ consists of functions $f:D \rightarrow \mathbb{R}$ whose partial derivatives through order $\lfloor\beta\rfloor$ exist and satisfy 
    \begin{align*}
         \sum_{\boldsymbol\alpha:1\le |\boldsymbol\alpha| < \beta} \Vert \partial^{\boldsymbol\alpha}f\Vert_\infty + \sum_{\boldsymbol\alpha:|\boldsymbol\alpha|  =\lfloor\beta\rfloor} \sup_{x,y \in D, x\neq y}\frac{|\partial^{\boldsymbol\alpha}f(x) - \partial^{\boldsymbol\alpha}f(y)|}{\Vert x-y\Vert^{\beta - \lfloor\beta\rfloor}_\infty} \le K.
    \end{align*}
\end{definition}

For $0 < \beta \le 1$, the first sum is empty, and the second term controls the Hölder continuity of $f$ itself. Thus, omitting its supremum norm does not remove smoothness control.

\begin{definition}[Unbounded Hölder compositional class]\label{def:unbdd_Holder_comp}
    Let $\boldsymbol{D}=(D_0,\ldots,D_{q+1})$, where each $D_i \subseteq \mathbb{R}^{d_i}$ is a Cartesian product of intervals containing the origin. The class $\mathcal{F}_\mathcal{D}(q,\boldsymbol{d},\boldsymbol{t},\beta,\boldsymbol{K};\boldsymbol{D})$ consists of compositions $f = f_q \circ\cdots\circ f_0$ with $f_i : D_i \to D_{i+1}$ such that, for every layer $i$ and coordinate $j$, there exist a fixed set $S_{i,j} \subseteq \{1,\ldots,d_i\}$, with $1 \le |S_{i,j}| \le t_i$, and a function $\tilde{f}_{i,j}$ such that
    \[
    f_{i,j}(z) = \tilde{f}_{i,j}(z_{S_{i,j}}),\quad \tilde{f}_{i,j} \in \mathcal{D}_{|S_{i,j}|}^{\beta_i}\left(\pi_{S_{i,j}}(D_i),K_i\right).
    \]
\end{definition}

The following example illustrates why allowing unbounded function values is necessary even for a simple Gaussian location model.

\begin{example}[Unbounded conditional support]\label{ex:ex3}
    Let $X \in [0,1]$ and let $\epsilon \sim N(0,1)$ be independent of $X$. The map $f(x,\epsilon)=x+\epsilon$ is Lipschitz with constant 2 under the maximum norm and satisfies
    \[
    f \in \mathcal{D}^1_2([0,1]\times\mathbb{R},2), \quad \Vert f\Vert_\infty=\infty.
    \]
    It therefore belongs to no $\mathcal{C}^1_2([0,1]\times\mathbb{R}, K)$ with finite $K$, while $Y=f(X,\epsilon)$ has conditional distribution $Y \mid X=x \sim N(x,1)$. The unbounded class thus permits unbounded conditional support while preserving smoothness.
\end{example}

\subsection{Finite-Sample Bounds under Gaussian Structural Noise}\label{sec:bound_Gaussian}

We specialize the structural representation in Assumption~\ref{assu:mixed_model} to the following model with shared Gaussian structural noise and impose compositional smoothness on its continuous and binary components. The continuous generator is defined on an unbounded structural-noise domain, whereas the latent binary score is defined on the uniform-noise cube.
For $a \in \{b,c\}$, indexing the binary and continuous components, let $\boldsymbol{d}^a = (d_{a,0},\ldots,d_{a,q_a+1})^\top$, $\boldsymbol{t}^a = (t_{a,0},\ldots,t_{a,q_a})^\top$, and use analogous indexing for $\boldsymbol{\beta}^a$ and $\boldsymbol{K}^a$. The input widths are $d_{b,0}=d_{c,0}=p+k$, and the output widths are $d_{b,q_b+1}=d_b$ and $d_{c,q_c+1}=d_c$.

\begin{assumption}[Shared Gaussian structural noise compositional model]\label{assu:mixed}
    There exists $\epsilon \sim N(0,I_k)$ independent of $X$, a continuous generator $f^c:[0,1]^p \times \mathbb{R}^k \to \mathbb{R}^{d_c}$, and a latent binary score $f^b: [0,1]^p \times [0,1]^k \to \mathbb{R}^{d_b}$, such that 
    \[
    Y^c = f^c(X,\epsilon),\qquad 
    Y^b = \mathbf{1}\left\{f^b(X,\Phi_0(\epsilon)) \ge 0\right\},
    \]
    where $\Phi_0$ is the standard normal distribution function applied coordinatewise. Moreover,
    \[
    f^c \in \mathcal{F}_{\mathcal{D}}(q_c,\boldsymbol{d}^c,\boldsymbol{t}^c,\boldsymbol\beta^c,\boldsymbol{K}^c;\boldsymbol{D}),\qquad
    f^b \in \mathcal{F}(q_b,\boldsymbol{d}^b,\boldsymbol{t}^b,\boldsymbol\beta^b,\boldsymbol{K}^b),
    \]
    where $\boldsymbol{D} = ([0,1]^p\times \mathbb{R}^k, \mathbb{R}^{d_{c,1}}, \ldots,\mathbb{R}^{d_{c,q_c+1}})$. 
    The continuous component satisfies the anchor bound
    \[
    \max_{0 \le i\le q_c}\max_{1\le j \le d_{c,i+1}}|f^c_{i,j}(0)| \le C_\mathrm{anc}
    \]
    for some positive constant $C_\mathrm{anc}$.
\end{assumption}

The continuous component receives $\epsilon$ directly, allowing unbounded conditional support. In the binary component, $U=\Phi_0(\epsilon) \sim \mathrm{Unif}((0,1)^k)$ places the latent score on a bounded domain. Both components depend on the same structural noise, so the representation continues to permit dependence within and across response blocks. Requiring $f^b$ to admit a smooth compositional representation on the closed uniform-noise cube is a structural restriction. The anchor bound controls the component values at the origin, which are not bounded by membership in $\mathcal{F}_\mathcal{D}$.

\begin{assumption}[Uniform threshold margin]\label{assu:reg_mixed}
    There exist constants $r_0 > 0$ and $C_\gamma < \infty$ such that, for $U \sim \mathrm{Unif}([0,1]^k)$ and each binary coordinate $j=1,\ldots,d_b$, 
    \[
    \sup_{x \in [0,1]^p} \mathrm{Pr}\left(|f^b_j(x,U)| \le t\right) \le C_\gamma t, \qquad 0<t\le r_0.
    \]
\end{assumption}

Assumption~\ref{assu:reg_mixed} limits, uniformly in $x$, the probability over $U$ that $f_j^b(x,U)$
lies near zero, thereby controlling the sensitivity
of thresholding to perturbations of the latent score.
It is analogous to margin conditions for plug-in classification
\citep{Audibert_2007}, which concern class probabilities
near $1/2$.
The following example verifies the uniform margin condition for a simple mixed-response model with shared Gaussian structural noise.

\begin{example}[Uniform threshold margin in a mixed-response model]\label{ex:ex4}
    Consider $X \in [0,1]$, $Y^c=X+\epsilon$ and $Y^b = \mathbf{1}\{\pi(X)-\Phi_0(\epsilon) \ge 0\}$, where $\epsilon \sim N(0,1)$ and $\pi:[0,1] \to [0,1]$ is smooth. The latent score $f^b(x,u)=\pi(x)-u$ satisfies
    \[
    \mathrm{Pr}\left(|f^b(x,U)|\le t\right) = \mathrm{Pr}\left(|U-\pi(x)| \le t\right) \le 2t.
    \]
    Thus the margin condition holds uniformly in $x$ with $C_\gamma=2$, while the shared structural noise links the continuous and binary responses.
\end{example}

We now establish nonasymptotic bounds for both the augmented error $\mathcal{E}_\Phi(\hat{g}_{n,m})$ of the fitted soft generator and the full-response error $\mathcal{E}_F(\mathcal{T}\hat{g}_{n,m})$ of the final thresholded sampler. The augmented error bound combines stochastic error from the observed data and Monte Carlo samples with approximation error from the sparse ReLU generator class. The threshold-transfer inequality then yields the corresponding bound for the final sampler.

\vspace{1mm}
\noindent\textbf{Rate Parameters.}
The error bound depends on the effective smoothness, the size of the fitted network, and the approximation exponents.
Let $a \in \{b,c\}$ index the binary and continuous components. For layer $i=0,\ldots,q_a$, define 
\[
\beta_{a,i}^* = \beta_{a,i}\prod_{\ell=i+1}^{q_a}(\beta_{a,\ell}\wedge 1), \quad \beta_a^* = \prod_{i=0}^{q_a}(\beta_{a,i}\wedge 1).
\]
Here, $\beta^*_{a,i}$ denotes the smoothness of layer $i$ after propagation through the subsequent layers, while $\beta_a^*$ aggregates the Hölder exponents of the entire composition.
Set
\[
\lambda_a = \max_{0 \le i \le q_a} \frac{t_{a,i}}{t_{a,i}+\beta_{a,i}}, \quad \gamma_a = \min_{0 \le i \le q_a}\frac{\beta^*_{a,i}}{t_{a,i}+\beta_{a,i}}, \quad \lambda = \max\left\{\frac{1}{2},\lambda_b,\lambda_c\right\}.
\]
Here, $\lambda_a$ reflects the largest layerwise network-size requirement, whereas $\gamma_a$ reflects the slowest approximation rate after error propagation through subsequent layers. The common exponent $\lambda$ determines network width and sparsity and thus enters the stochastic bound, while $\gamma_b$ and $\gamma_c$ contribute to the approximation exponent $r$ defined below.
For the Gaussian approximation term, define
\[
\gamma = \min\left\{ \frac{\beta^*_b\beta^*_c}{2},\beta^*_c\gamma_b, \gamma_c \right\}, \quad \eta = 1+\sum_{i=1}^{q_c}\beta_{c,i}, \quad r = \frac{\gamma}{1+2\beta_{c,0}+\beta_{b,0}+\max\{ \beta_{b,0}^*\beta_c^*, 2\beta_{c,0}^* \}},
\]
where $r$ is the polynomial approximation exponent, whereas $\eta$ enters only through a logarithmic factor in the error bound. With fixed positive smoothness indices and finite intrinsic dimensions, $\lambda<1$ and $r>0$.

For $n \ge 3$, let
\[
\ell_n(p,d,k) = \log n\log(p+k) + (\log n)^2\log d + (\log n)^3
\]
and define
\begin{align}\label{eq:nonasymp_bound_Gaussian}
    \mathcal{S}_{n,m}(\delta) &= \left( \sqrt{\frac{1}{n}}+\sqrt{\frac{1}{nm}} \right) \left[dn^\lambda\ell_n(p,d,k) + \log\left(\frac{1}{\delta}\right)\right]^{1/2},\notag\\
    B_{n,m,\delta} &= \mathcal{S}_{n,m}(\delta)+ \left\{ \sqrt{d}(\log n)^{\eta/2} + k(\log n)^{-1/2}\right\} n^{-r}.
\end{align}

In Theorems~\ref{thm:bound_mixed} and \ref{thm:bound_t}, the constants implicit in \(\lesssim\) depend only on \(\omega\) and the compositional parameters \((q_a,\mathbf t_a,\boldsymbol\beta_a,\mathbf K_a)\), \(a\in\{b,c\}\). They are independent of \(n,m,\delta\), the ambient dimensions \(p,d,k\), and the intermediate structural widths.

\begin{theorem}[Finite-sample bounds under Gaussian structural noise]\label{thm:bound_mixed}
    Suppose Assumptions~\ref{assu:mixed} and \ref{assu:reg_mixed} hold, and let $m \ge 2$. Choose the sparse ReLU architecture with $L \asymp \log n$ and $\boldsymbol{p}=(p+k,p_1,\ldots,p_L,d)$, where $\min_{\ell=1,\ldots,L}p_\ell \asymp dn^\lambda$ and $s \asymp dn^\lambda\log n$.
    Then, for every $\delta \in (0,1)$ and sufficiently large $n$, with probability at least $1-\delta$, both
\begin{align*}
    \mathcal{E}_\Phi(\hat{g}_{n,m}) = \mathbb{E}_X[\mathrm{ED}_\Phi(P_{\hat{g}_{n,m}\mid X},P_{Y \mid X})] &\lesssim B_{n,m,\delta},
\end{align*}
and 
    \[
    \mathcal{E}_F(\mathcal{T}\hat{g}_{n,m}) \le \frac{2}{1-\omega}\mathcal{E}_\Phi(\hat{g}_{n,m}) + \frac{25}{4\sqrt{\omega}}\{\mathcal{E}_\Phi(\hat{g}_{n,m})\}^{1/2} \lesssim B_{n,m,\delta}+B_{n,m,\delta}^{1/2},
    \]
    hold on the same event.
\end{theorem}

The stochastic term $S_{n,m}(\delta)$ accounts for the observed sample and the Monte Carlo approximation of the loss, while the remaining term in \eqref{eq:nonasymp_bound_Gaussian} measures the approximation of the structural model by the fitted generator class. The polynomial exponents depend on the compositional structure and smoothness, and the bound explicitly tracks dependence on the ambient dimensions.

In particular, when \(p,d,k\) and the parameters in the structural assumptions are fixed, Theorem~\ref{thm:bound_mixed} yields
\[
\begin{aligned}
\mathcal E_\Phi(\hat g_{n,m}) &= O_p\!\left( n^{-(1-\lambda)/2}(\log n)^{3/2} + n^{-r}(\log n)^{\eta/2} \right),\\
\mathcal E_F(\mathcal{T}\hat g_{n,m}) &= O_p\!\left( n^{-(1-\lambda)/4}(\log n)^{3/4} + n^{-r/2}(\log n)^{\eta/4} \right).
\end{aligned}
\]
Thus, up to logarithmic factors, the upper bound for the final thresholded full-response error is of order \(n^{-\min\{(1-\lambda)/4,\,r/2\}}\). Since \(\lambda<1\) and \(r>0\), both errors converge to zero in probability, even when the number of auxiliary-noise draws per observation is held fixed at \(m\ge2\).

\subsection{Mixed Responses with Student-$t$ Structural Noise}\label{sec:bound_t}

To accommodate heavy-tailed continuous responses, we extend the mixed-response analysis to structural generators driven by shared Student-$t$ structural noise.
We retain the augmented discrepancy $\Phi$, the clipped generator class $\mathcal{G}_\text{fin}$, and the empirical estimator defined in Section~\ref{sec:sample_gen_engression}. The fitted generator still uses uniform auxiliary noise. Throughout this subsection, $\nu > 0$ is fixed, and $F_\nu$ denotes the univariate Student-$t_\nu$ distribution function, applied coordinatewise.

\begin{assumption}[Shared Student-$t$ compositional model]\label{assu:mixed_t}
    There exists $\epsilon=(\epsilon_1,\ldots,\epsilon_k)^\top$, independent of $X$, with $\epsilon_j\overset{\text{i.i.d.}}{\sim} t_\nu$, such that
    \[
    Y^c = f^c(X,\epsilon), \quad Y^b = \mathbf{1}\left\{f^b(X,F_\nu(\epsilon)) \ge 0\right\}.
    \]
    The functions $f^c,f^b$ satisfy the same domain, compositional class, and anchor requirements as in Assumption~\ref{assu:mixed}. The latent binary score satisfies Assumption~\ref{assu:reg_mixed} with $U \sim \mathrm{Unif}([0,1]^k).$
\end{assumption}

Retain $\lambda, \gamma_b,\gamma_c$ and the effective smoothness quantities from Section~\ref{sec:bound_Gaussian}. Define
\[
\Gamma_c = \min\{\beta_c^*/2, \gamma_c\}
\]
and the continuous radius-growth exponent
\[
\eta_{c,\nu} = 1+(\nu+1)\beta_{c,0} + \sum_{i=1}^{q_c}\beta_{c,i} + \max \left\{ (\nu+1)\beta_{c,0}^*, \max_{1 \le i \le q_c-1} \beta^*_{c,i}, 1+\beta^*_{c,q_c} \right\}.
\]
The maximum over $1 \le i \le q_c-1$ is omitted when its index set is empty. Set
\[
r_{c,\nu} = \frac{\nu\Gamma_c}{\nu+\eta_{c,\nu}}, \quad r_\nu = \min\{\gamma_b, r_{c,\nu}\}, \quad B_{n,m,\delta}^{(\nu)} = S_{n,m}(\delta) + (\sqrt{d}+k)n^{-r_\nu}.
\]
Here, $r_\nu$ combines the binary approximation exponent $\gamma_b$ and the continuous approximation exponent $r_{c,\nu}$, which depends on the Student-$t$ structural noise.

\begin{theorem}[Finite-sample bounds under Student-$t$ structural noise]\label{thm:bound_t}
    Suppose Assumption~\ref{assu:mixed_t} holds, and let $m \ge 2$. Choose the sparse ReLU architecture with $L \asymp \log n$ and $\boldsymbol{p}=(p+k,p_1,\ldots,p_L,d)$, where $\min_{\ell=1,\ldots,L}p_\ell \asymp dn^\lambda$ and $s \asymp dn^\lambda\log n$.
    Then, for every $\delta \in (0,1)$ and sufficiently large $n$, with probability at least $1-\delta$, both
\begin{align*}
    \mathcal{E}_\Phi(\hat{g}_{n,m}) = \mathbb{E}_X[\mathrm{ED}_\Phi(P_{\hat{g}_{n,m}\mid X},P_{Y \mid X})] &\lesssim B_{n,m,\delta}^{(\nu)},
\end{align*}
and 
    \[
    \mathcal{E}_F(\mathcal{T}\hat{g}_{n,m}) \le \frac{2}{1-\omega}\mathcal{E}_\Phi(\hat{g}_{n,m}) + \frac{25}{4\sqrt{\omega}}\{\mathcal{E}_\Phi(\hat{g}_{n,m})\}^{1/2} \lesssim B_{n,m,\delta}^{(\nu)}+\{B_{n,m,\delta}^{(\nu)}\}^{1/2},
    \]
    hold on the same event.
\end{theorem}

The stochastic term $S_{n,m}(\delta)$ has the same form as in Theorem~\ref{thm:bound_mixed} because the loss remains bounded and the fitted network class has the same size. The degrees of freedom $\nu$ enter the continuous approximation exponent $r_{c,\nu}$ and hence the overall approximation exponent $r_\nu$. In the fixed-dimensional setting, with $p,d,k$, the structural regularity parameters, and $\nu>0$ held fixed, Theorem~\ref{thm:bound_t} yields
\[
\mathcal{E}_F(\mathcal{T}\hat{g}_{n,m}) = O_p\left( n^{-(1-\lambda)/4}(\log n)^{3/4} + n^{-r_\nu/2} \right).
\]
Since $\lambda<1$ and $r_\nu>0$ for every fixed $\nu>0$, the thresholded full-response error converges to zero in probability as $n\to\infty$. Thus, the final thresholded sampler is consistent under the Student-$t$ structural model, even when the number of auxiliary-noise draws per observation is held fixed at $m\ge2$.

\subsection{High-Dimensional Consistency of the Thresholded Generator}\label{sec:consistency}

We study a high-dimensional setting to determine how rapidly the covariate, response, and structural-noise dimensions may grow with the sample size while retaining consistency of the final thresholded sampler. The conditions below allow the ambient dimensions to increase while keeping the smoothness and intrinsic dimensions uniformly controlled.

Consider a sequence of mixed-response models with covariates $X_n \in [0,1]^{p_n}$ and responses 
\[
Y_n = (Y^c_n,Y^b_n) \in \mathbb{R}^{d_{c,n}}\times \{0,1\}^{d_{b,n}}, \quad d_n = d_{c,n}+d_{b,n}.
\]
The structural maps $f_n^c$ and $f^b_n$ use $k_n$-dimensional structural noise and follow either the Gaussian representation in Assumption~\ref{assu:mixed} or the Student-$t$ representation in Assumption~\ref{assu:mixed_t}.
The same noise family is used throughout the sequence. Write $F=\Phi_0$ in the Gaussian case and $F=F_\nu$ in the Student-$t$ case, with fixed $\nu >0$. In both cases, the fitted generator continues to use uniform auxiliary noise.

To make the changing response dimensions explicit, write $\hat{g}_n = \hat{g}_{n,m_n}$, and let $\mathrm{ED}_{F,n}$ and $\mathrm{ED}_{\Phi,n}$ denote the energy distances in Definition~\ref{def:mixed_dis} with block dimensions $(d_{c,n},d_{b,n})$. Define
\begin{align*}
    \mathcal{E}_{\Phi,n}(g) = \mathbb{E}_{X_n}[\mathrm{ED}_{\Phi,n}(P_{g \mid X_n},P_{Y_n \mid X_n})], \quad
    \mathcal{E}_{F,n}(\mathcal{T}g) = \mathbb{E}_{X_n}[\mathrm{ED}_{F,n}(P_{\mathcal{T}g \mid X_n},P_{Y_n \mid X_n})].
\end{align*}
The input and output widths of the structural compositions satisfy $d^a_{n,0}=p_n+k_n$ and $d^a_{n,q_a+1}=d_{a,n}$ for $a \in \{b,c\}$; their intermediate widths may vary with $n$.

\begin{assumption}[Uniform regularity along the sequence]\label{assu:uniform_reg}
    For each $a \in \{b,c\}$, the compositional depth $q_a$, intrinsic-dimension vector $\boldsymbol{t}^a$, smoothness vector $\boldsymbol{\beta}^a$, and radius vector $\boldsymbol{K}^a$ do not depend on $n$. For every $n$,
    \[
    f^c_n \in \mathcal{F}_\mathcal{D}(q_c,\boldsymbol{d}_n^c,\boldsymbol{t}^c,\boldsymbol{\beta}^c,\boldsymbol{K}^c;\mathcal{D}_n^c), \quad f^b_n \in \mathcal{F}(q_b,\boldsymbol{d}_n^b,\boldsymbol{t}^b,\boldsymbol{\beta}^b,\boldsymbol{K}^b),
    \]
    with the component domains specified in Assumption~\ref{assu:mixed}, using the $n$th dimensions. There is a finite constant $C_\mathrm{anc}$ such that
    \[
    \sup_{n \ge 1}\max_{0 \le i \le q_c} \max_{1 \le j \le d_{n,i+1}^c} |f^c_{n,i,j}(0)| \le C_\mathrm{anc}.
    \]
    The margin constants $r_0,C_\gamma$ in Assumption~\ref{assu:reg_mixed} are uniform in $n$ and across the binary coordinates, and $\omega \in (0,1)$ is fixed.
\end{assumption}

The structural conditions keep local compositional complexity fixed while allowing the ambient dimensions and intermediate widths to grow. The selected input coordinates may change with $n$, but their number remains bounded by $t_{a,i}$, and their selection does not depend on the input value. 
The exponents \(\lambda,r,\eta\), and \(r_\nu\), as applicable, are consequently fixed along the sequence, and the implicit constants in Theorems~\ref{thm:bound_mixed} and \ref{thm:bound_t} can be chosen uniformly in \(n\).
The following corollary gives sufficient growth conditions on \(p_n\), \(d_n\), and \(k_n\) for polynomial convergence rates under either structural-noise model.

\begin{corollary}[High-dimensional consistency under Gaussian or Student-$t$ structural noise]\label{cor:consistency}
    Suppose the sequence above satisfies Assumption~\ref{assu:uniform_reg}. For each $n$, choose the sparse ReLU architecture with $L_n \asymp \log n$ and $\boldsymbol{p}_n=(p_n+k_n,p_{n,1},\ldots,p_{n,L_n},d_n)$, where $\min_{\ell=1,\ldots,L_n}p_{n, \ell} \asymp d_nn^\lambda$ and $s \asymp d_nn^\lambda\log n$, with the $n$th dimensions and comparison constants uniform in $n$, and let $m_n \ge 2$. Set
    \begin{align*}
        r_* = \begin{cases}
            r, &F=\Phi_0 \quad \text{(Gaussian noise)},\\
            r_\nu, &F=F_\nu \quad \text{(Student-$t_\nu$ noise, fixed $\nu > 0$)}.
        \end{cases}
    \end{align*}
    Fix $0 < \zeta < \min\{(1-\lambda)/4,r_*/2\}$, and choose $\delta_n \downarrow 0$ with $\log(1/\delta_n) \lesssim n^{1-4\zeta}$. Suppose that, for constants $a_p,a_d,a_k \ge 0$,
    \[
    d_n \lesssim n^{a_d}, \quad k_n \lesssim n^{a_k}, \quad p_n \lesssim \exp(n^{a_p}),
    \]
    and 
    \[
    a_d+a_p < 1-\lambda-4\zeta, \quad a_d < 2(r_*-2\zeta), \quad a_k < r_*-2\zeta.
    \]
    Then
    \[
    \mathcal{E}_{\Phi,n}(\hat{g}_n) = O_p(n^{-2\zeta}), \quad \mathcal{E}_{F,n}(\mathcal{T}\hat{g}_n) = O_p(n^{-\zeta}).
    \]
    In particular, $\mathcal{E}_{F,n}(\mathcal{T}\hat{g}_n) \overset{p}{\to} 0$.
\end{corollary}

For Gaussian structural noise, the dimension conditions use the approximation exponent $r$. For Student-$t_\nu$ structural noise, they use $r_\nu$ instead, so the allowable growth rates of the response and noise dimensions depend on the tail parameter through the approximation exponent.

\section{Numerical Study}\label{sec:simulation}

In this section, we evaluate mixed engression for conditional distributions with continuous and binary responses. The experiments examine sample-size behavior, the effects of the full-response energy loss and binary augmentation, and sensitivity to the structural noise distribution. The primary evaluation measure is the bounded full-response energy distance $\mathrm{ED}_F$, which assesses the final mixed-response sampler. The Euclidean energy distance $\mathrm{ED}$ and the binary energy distance $\mathrm{ED}_B$ provide  complementary assessments of the full mixed distribution and its binary component, respectively.

\subsection{Experimental Design}\label{sec:dgp}

\textbf{Data-generating model.}
We generate $q$ continuous and $q$ binary responses, giving a total response dimension of $d=2q$. The covariates $X_1,\ldots,X_p$ have independent $\mathrm{Unif}(0,1)$ coordinates, and their effect is summarized by 
\[
s(X) = \sqrt{\frac{3}{p}}\sum_{\ell=1}^p (2X_\ell-1), \quad \mathbb{E} [s(X)] = 0, \quad \mathrm{Var}[s(X)] = 1.
\]
The latent scores and observed responses are
\[
H = s(X)\mathbf{1}_d + c+A\epsilon, \quad Y_j^c = H_{2j-1}, \quad Y^b_j = \mathbf{1}\{H_{2j}>0\}, \quad j=1,\ldots,q.
\]
The coordinate-specific offsets are $c_j=(j-1)/(d-1)-1/2$, for $j=1,\ldots,d$. To construct $A$, let $C_\rho$ have entries $(C_\rho)_{ab}=\rho^{|a-b|}$ with $\rho=0.8$, and let $L$ be its lower-triangular Cholesky factor.
We set $A=LP$, where $P$ is the orthogonal factor obtained from the QR decomposition of a Gaussian random matrix generated with seed 2026; consequently, $AA^\top=C_\rho$. The standardized shared index $s(X)$ keeps the covariate-signal variance fixed across values of $p$, while the shared noise term $A\epsilon$ induces dependence within and across the response blocks.

The main experiments use $p \in \{5,10\}$ and $q \in \{3,5,10\}$, with $n \in \{2000,5000,10000\}$ training observations. We set the auxiliary noise dimension to $k=2q$ and use independent standard Gaussian or Student-$t_5$ coordinates for $\epsilon$. At $n=10000$ and $p=q=10$, we additionally consider $t_3$ and $t_2$ structural noise distributions. Within each configuration, the response-generating mechanism is held fixed as $n$ varies. The results reported below focus on $p=10$ and $q \in \{3,10\}$.

\vspace{1mm}
\noindent\textbf{Methods and fitting.}
We compare mixed engression with $\omega \in \{0,0.05,0.1,0.2\}$, standard engression, distributional random forests (DRF) \citep{JMLR:v23:21-0585}, a 50-nearest-neighbor estimator (kNN), and a combined kNN and logistic-regression estimator (kNN+logit). The kNN+logit estimator resamples the continuous block from the 50 nearest neighbors and generates the binary coordinates from separate logistic regressions, independently of the continuous block given $X$. 

Within each repetition, all engression methods use the same training data, network architecture, initialization, and optimization schedule. Mixed engression with $\omega=0$ uses only the full-response energy loss, whereas positive weights use $(1-\omega)\varphi_F + \omega\varphi_B$, thereby adding the binary discrepancy. 
Shared neural network settings are summarized in Table~\ref{tab:simulation_settings}.

\begin{table}[!t]
\centering
\caption{Shared neural network settings.}
\label{tab:simulation_settings}
\small
\renewcommand{\arraystretch}{1.14}
\setlength{\tabcolsep}{5pt}

\begin{tabularx}{\linewidth}{@{}p{0.25\linewidth}X@{}}
\toprule
Setting & Values \\

\midrule

Generator
& Fully connected network with three hidden layers, each containing 100 units with ReLU activation \\

Auxiliary noise
& $\xi \sim \operatorname{Unif}([0,1]^k)$,
  $k=2q$ \\

Optimization
& Adam; learning rate $5 \times 10^{-5}$;
  1,000 epochs; batch size 256 \\

Training draws
& 200 auxiliary noise draws per observation/update \\

\bottomrule
\end{tabularx}
\end{table}

\vspace{1mm}
\noindent\textbf{Evaluation.}
For a fitted conditional distribution $\hat{P}_{Y \mid X}$, we assess
\[
\mathbb{E}_X[\mathrm{ED}_F(\hat{P}_{Y \mid X}, P_{Y \mid X})], \quad \mathbb{E}_X[\mathrm{ED}(\hat{P}_{Y \mid X}, P_{Y \mid X})], \quad \mathbb{E}_X[\mathrm{ED}_B(\hat{P}_{Y^b \mid X}, P_{Y^b \mid X})].
\]
We refer to these quantities as $\mathrm{ED}_F$, $\mathrm{ED}$, and $\mathrm{ED}_B$ in the results. Both $\mathrm{ED}_F$ and $\mathrm{ED}$ assess the full mixed conditional distribution, whereas $\mathrm{ED}_B$ assesses the conditional distribution of the entire binary vector, including dependence among the binary coordinates. 

Each repetition uses $N=5000$ test covariates drawn from $\mathrm{Unif}([0,1]^p)$. At each test covariate $x_i$, we generate $M=500$ draws $Y_{i1},\ldots,Y_{iM}$ from the true conditional distribution and $M$ independent draws $Z_{i1},\ldots,Z_{iM}$ from the fitted conditional distribution.
For a discrepancy $\varphi$, the following estimator is unbiased for $\mathrm{ED}_\varphi\left(\hat{P}_{Y \mid X=x_i},P_{Y \mid X=x_i}\right)$, conditional on the fitted model and $x_i$:
\[
\hat{D}_{\varphi,i} = \frac{2}{M}\sum_{j=1}^M \varphi(Y_{ij},Z_{ij}) - \frac{1}{M(M-1)} \sum_{j \neq \ell} \{\varphi(Y_{ij},Y_{i,\ell}) + \varphi(Z_{ij},Z_{i\ell})\}.
\]
For each metric, the repetition-level estimate is $N^{-1}\sum_{i=1}^N \hat{D}_{\varphi,i}$.
Test covariates and true conditional samples are shared across methods. We report means and standard errors of these repetition-level estimates across 20 repetitions.

\subsection{Simulation Results}\label{sec:sim_result}

\textbf{Sample-size behavior and overall performance}

\begin{figure}[!t]
    \centering
    \includegraphics[width=\textwidth]{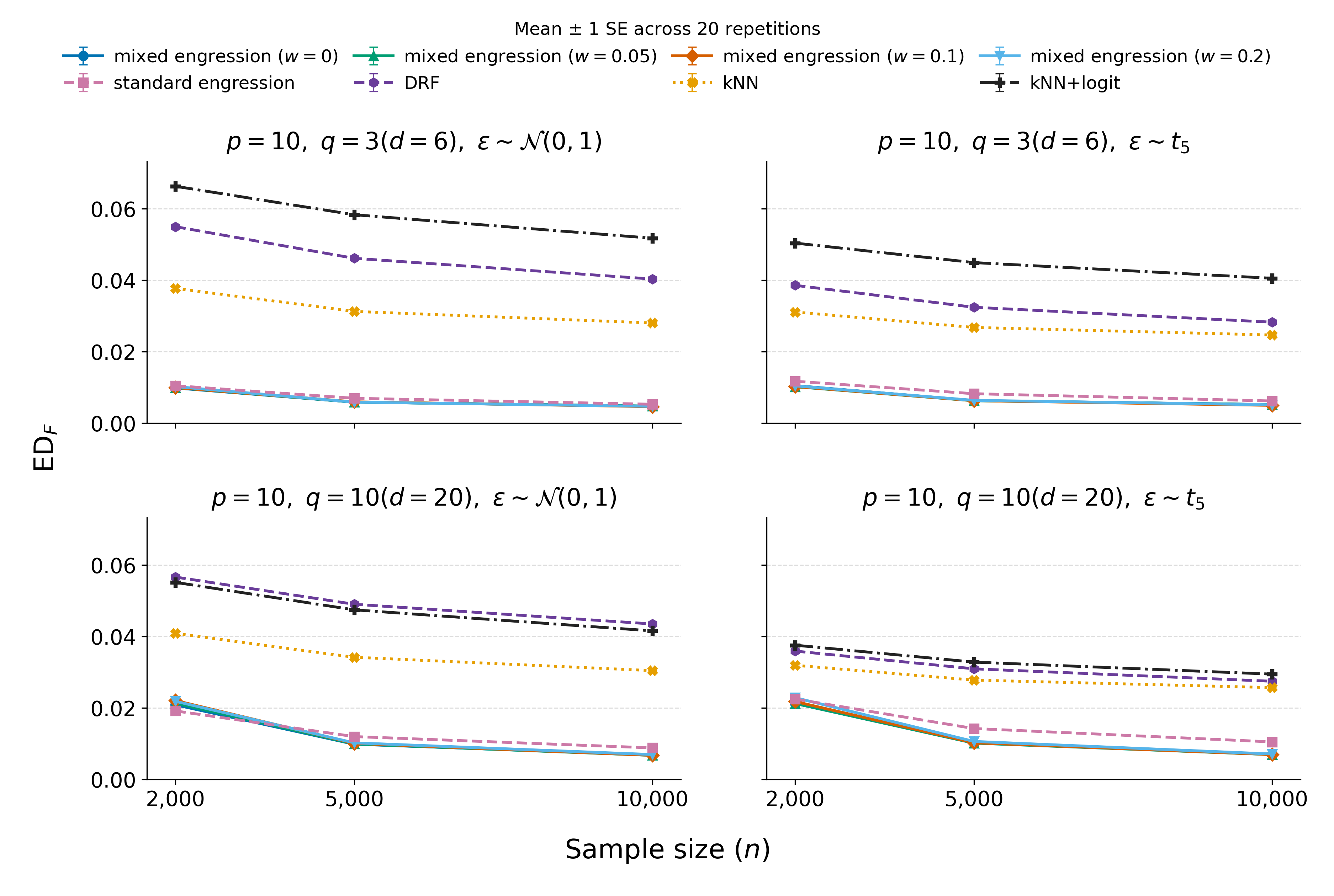}
    \caption{Sample-size behavior of $\mathrm{ED}_F$ for mixed-response conditional distribution estimation. Subfigures use $p=10$ and $(q,\text{noise}) = (3,N(0,1)),(3,t_5),(10,N(0,1)),(10,t_5)$, respectively. Points and error bars show means and one standard error over 20 repetitions at $n=2000,5000,10000$.}
    \label{fig:sim_sample_size}
\end{figure}

\vspace{1mm}
Figure~\ref{fig:sim_sample_size} compares $\mathrm{ED}_F$ across training sample sizes for $p =10$, $q \in \{3,10\}$, and Gaussian or $t_5$ structural noise. Mean $\mathrm{ED}_F$ values decrease with $n$ for every method displayed. The engression methods attain lower mean energy distances than DRF, kNN, and kNN+logit at every displayed sample size. These patterns indicate improvement with additional training data and a persistent advantage of the neural conditional generators under the full-response energy distance in these configurations.

The common vertical scale accommodates the larger energy distances of the baseline methods and compresses differences among the engression methods. In particular, the relative performance of standard and mixed engression at $n=10000$, as well as the smaller differences among augmentation weights, is difficult to assess from the curves. We therefore use Table~\ref{tab:engression-ed} for these comparisons. Its first four configurations match those in Figure~\ref{fig:sim_sample_size}, while the remaining two add $t_3$ and $t_2$ noise.

\vspace{3mm}
\noindent\textbf{Effects of the full-response energy loss and binary augmentation}

\begin{table}[!t]
\centering
\caption{Energy distances for standard and mixed engression at $n=10000$. Row labels specify $(p,q,\text{noise})$, where $N$ denotes standard Gaussian noise. Blocks A, B, and C assess $\mathrm{ED}_F$, $\mathrm{ED}$, and $\mathrm{ED}_B$, respectively. Entries are $1000 \times (\text{mean}\pm\text{SE})$ over 20 repetitions, and boldface marks the smallest mean in each row.}
\label{tab:engression-ed}
\resizebox{\textwidth}{!}{%
\begin{tabular}{lccccc}
\toprule
& Standard & \multicolumn{4}{c}{Mixed engression} \\
\cmidrule(lr){3-6}
$(p,q,\mathrm{noise})$
& engression
& $\omega=0$
& $\omega=0.05$
& $\omega=0.1$
& $\omega=0.2$ \\
\midrule
\multicolumn{6}{l}{\textbf{Block A: $\mathrm{ED}_{F}$}} \\
\addlinespace
$(10,3,N)$
& $5.31 \pm 0.10$
& $4.77 \pm 0.09$
& $4.79 \pm 0.09$
& $\mathbf{4.65} \pm 0.10$
& $4.73 \pm 0.08$ \\
$(10,3,t_5)$
& $6.23 \pm 0.11$
& $5.23 \pm 0.09$
& $5.22 \pm 0.08$
& $\mathbf{5.01} \pm 0.09$
& $5.16 \pm 0.08$ \\
$(10,10,N)$
& $8.84 \pm 0.21$
& $6.87 \pm 0.14$
& $\mathbf{6.75} \pm 0.14$
& $6.76 \pm 0.16$
& $6.99 \pm 0.13$ \\
$(10,10,t_5)$
& $10.52 \pm 0.28$
& $7.09 \pm 0.16$
& $\mathbf{6.97} \pm 0.12$
& $6.98 \pm 0.16$
& $7.14 \pm 0.18$ \\
$(10,10,t_3)$
& $12.41 \pm 0.32$
& $\mathbf{7.18} \pm 0.15$
& $7.20 \pm 0.13$
& $7.18 \pm 0.10$
& $7.61 \pm 0.16$ \\
$(10,10,t_2)$
& $15.14 \pm 0.46$
& $\mathbf{7.46} \pm 0.15$
& $7.50 \pm 0.17$
& $7.55 \pm 0.12$
& $8.37 \pm 0.15$ \\

\midrule
\multicolumn{6}{l}{\textbf{Block B: $\mathrm{ED}$}} \\
\addlinespace
$(10,3,N)$
& $12.18 \pm 0.28$
& $11.51 \pm 0.26$
& $11.34 \pm 0.31$
& $11.18 \pm 0.26$
& $\mathbf{10.98} \pm 0.20$ \\
$(10,3,t_5)$
& $14.98 \pm 0.34$
& $13.61 \pm 0.32$
& $13.46 \pm 0.30$
& $13.10 \pm 0.30$
& $\mathbf{12.92} \pm 0.23$ \\
$(10,10,N)$
& $15.16 \pm 0.28$
& $13.89 \pm 0.61$
& $13.21 \pm 0.44$
& $12.81 \pm 0.38$
& $\mathbf{12.68} \pm 0.24$ \\
$(10,10,t_5)$
& $19.38 \pm 0.39$
& $21.31 \pm 2.56$
& $17.61 \pm 1.16$
& $17.30 \pm 1.26$
& $\mathbf{15.60} \pm 0.53$ \\
$(10,10,t_3)$
& $\mathbf{26.12} \pm 0.61$
& $43.99 \pm 7.42$
& $38.72 \pm 7.17$
& $34.84 \pm 5.18$
& $28.69 \pm 3.08$ \\
$(10,10,t_2)$
& $\mathbf{46.10} \pm 0.99$
& $204.65 \pm 32.64$
& $195.94 \pm 39.81$
& $159.86 \pm 27.30$
& $129.44 \pm 27.81$ \\

\midrule
\multicolumn{6}{l}{\textbf{Block C: $\mathrm{ED}_B$}} \\
\addlinespace
$(10,3,N)$
& $3.48 \pm 0.15$
& $3.00 \pm 0.11$
& $2.80 \pm 0.11$
& $2.64 \pm 0.12$
& $\mathbf{2.39} \pm 0.09$ \\
$(10,3,t_5)$
& $4.05 \pm 0.16$
& $3.36 \pm 0.14$
& $3.09 \pm 0.14$
& $2.73 \pm 0.13$
& $\mathbf{2.55} \pm 0.10$ \\
$(10,10,N)$
& $5.11 \pm 0.25$
& $4.14 \pm 0.14$
& $3.45 \pm 0.09$
& $3.31 \pm 0.08$
& $\mathbf{2.99} \pm 0.08$ \\
$(10,10,t_5)$
& $5.93 \pm 0.41$
& $4.50 \pm 0.17$
& $3.40 \pm 0.13$
& $3.10 \pm 0.12$
& $\mathbf{2.79} \pm 0.08$ \\
$(10,10,t_3)$
& $7.80 \pm 0.69$
& $5.40 \pm 0.34$
& $3.31 \pm 0.15$
& $2.89 \pm 0.08$
& $\mathbf{2.75} \pm 0.07$ \\
$(10,10,t_2)$
& $8.94 \pm 1.07$
& $9.08 \pm 0.47$
& $3.07 \pm 0.11$
& $2.87 \pm 0.07$
& $\mathbf{2.87} \pm 0.07$ \\

\bottomrule
\end{tabular}%
}
\end{table}

\vspace{1mm}
Table~\ref{tab:engression-ed} separates the contributions of the two components of the proposed loss. Comparing standard engression with mixed engression at $\omega=0$ assesses the effect of replacing the Euclidean energy loss with the full-response energy loss, while comparisons among the mixed engression methods assess the additional contribution of binary augmentation.

Block A shows that most of the improvement in $\mathrm{ED}_F$ is already achieved without binary augmentation. Mixed engression with $\omega=0$ attains a lower mean $\mathrm{ED}_F$ than standard engression in all six configurations. Positive augmentation weights provide smaller additional improvements, with reductions of approximately $1.6\%$--$4.3\%$ relative to $\omega=0$ for the best positive weights under Gaussian and $t_5$ noise. Thus, the full-response energy loss accounts for most of the improvement under the primary evaluation measure.

Binary augmentation has a more pronounced effect on the binary distribution, as shown in Block C. Every positive weight considered yields a lower mean $\mathrm{ED}_B$ than both standard engression and mixed engression with $\omega=0$ across all configurations. For example, under $t_2$ noise, increasing $\omega$ from $0$ to $0.1$ reduces $\mathrm{ED}_B$ from $9.08$ to $2.87$ on the scale reported in Table~\ref{tab:engression-ed}, a reduction of $68.4\%$. This improvement reflects the additional control provided by the binary energy distance, which directly compares the binary distributions without involving the continuous coordinates.

The augmentation weight also introduces a tradeoff between full-response and binary accuracy. Under $t_3$ and $t_2$ noise, the mean $\mathrm{ED}_F$ values for $\omega\in\{0,0.05,0.1\}$ are similar, whereas increasing the weight to $\omega=0.2$ worsens $\mathrm{ED}_F$. In particular, under $t_2$ noise, increasing $\omega$ from $0.1$ to $0.2$ leaves the mean $\mathrm{ED}_B$ essentially unchanged, while increasing $\mathrm{ED}_F$ from $7.55$ to $8.37$. These results indicate that moderate binary augmentation can substantially improve binary-distribution accuracy while largely preserving full-response accuracy, whereas a larger augmentation weight need not improve the overall fit.

\vspace{3mm}
\noindent\textbf{Sensitivity to the structural noise distribution}

\vspace{1mm}

For $p=q=10$, Table~\ref{tab:engression-ed} also allows us to examine sensitivity to the structural noise distribution across Gaussian, $t_5$, $t_3$, and $t_2$ noise. Block A shows that mixed engression is substantially more stable than standard engression under the full-response energy distance. As the structural noise changes from Gaussian to $t_2$, the mean $\mathrm{ED}_F$ increases by $71.2\%$ for standard engression, compared with only $8.6\%$ for mixed engression with $\omega=0$ and $11.7\%$ for mixed engression with $\omega=0.1$. Consequently, the relative advantage of mixed engression under $\mathrm{ED}_F$ becomes more pronounced in the heavier-tailed settings considered. A similar contrast appears in Block C: the mean $\mathrm{ED}_B$ for $\omega=0.1$ changes from $3.31$ to $2.87$, whereas the corresponding value for standard engression increases from $5.11$ to $8.94$.

However, the comparison under the Euclidean energy distance yields a different phenomenon. As shown in Block B, mixed engression generally improves on standard engression under Gaussian and $t_5$ structural noise, although mixed engression with $\omega=0$ performs worse at $q=10$ under $t_5$ noise. Under $t_3$ and $t_2$ noise, standard engression instead achieves the lowest mean Euclidean energy distance among the engression methods. Positive binary augmentation improves the Euclidean energy distance relative to $\omega=0$ in these settings but does not reverse the ordering.

These contrasting results are consistent with the different treatment of large pairwise separations by the two criteria. The bounded transformation $\rho(t)=t/(1+t)$ limits the contribution of large continuous differences, whereas the Euclidean discrepancy remains sensitive to their magnitude. As a result, the two energy distances can favor different fitted conditional distributions, particularly under heavy-tailed structural noise. The experiments therefore demonstrate improved accuracy under the proposed full-response energy distance and greater stability across the structural noise distributions considered, but not uniformly better performance under the Euclidean energy distance.

\section{Discussion}\label{sec:discussion}

In this paper, we propose mixed engression for estimating the conditional distribution of multivariate responses with continuous and binary components. The method aims to generate samples that respect the mixed support and approximate the full conditional distribution, including dependence within and across response types. It fits a continuous soft generator using an augmented energy loss that combines bounded full-response and binary energy losses and then thresholds the soft binary outputs for sampling. A central implication is that population identification and control after thresholding are distinct requirements. The full-response energy distance identifies the target distribution at each fixed response dimension, but this alone does not provide uniform control of the final sampler as the dimensions grow. Binary augmentation addresses this gap through a threshold-transfer bound: vanishing augmented error of the fitted soft generator implies vanishing full-response error of the thresholded sampler. Because both loss components are bounded, the population objective remains well defined without response-moment assumptions.

Our nonasymptotic analysis provides high-probability bounds for both the augmented error of the soft estimator and the full-response error of the thresholded sampler, with both errors averaged over the covariate distribution. Under compositional smoothness and a uniform threshold-margin condition, these guarantees apply to sparse ReLU estimators for mixed-response models driven by shared Gaussian or Student-\(t_\nu\) structural noise and allow unbounded continuous conditional support. For fixed dimensions and structural parameters, the bounds yield polynomial convergence rates in sample size, up to logarithmic factors. The Student-\(t\) results cover every fixed \(\nu>0\), including models without finite first response moments. The stochastic error term has the same form under both noise families, while the tail parameter affects the approximation rate. Under the stated uniform regularity and growth conditions, the bounds also yield polynomial convergence when the covariate, response, and noise dimensions increase while the local compositional complexity remains controlled.

The numerical results illustrate both the benefit of this construction and its dependence on the evaluation measure. In the heavier-tailed simulation settings considered, mixed engression achieves larger improvements under the full-response energy distance, while binary augmentation improves recovery of the binary component as measured by the binary energy distance. Standard engression can nevertheless perform better under the Euclidean energy distance. These findings are consistent with the different ways in which the discrepancies weight large pairwise distances: the bounded full-response discrepancy limits their contribution, whereas the Euclidean discrepancy remains sensitive to their magnitude. The binary evaluation further shows that a lower Euclidean energy distance can coexist with less accurate binary fit. The results therefore support the proposed loss when accuracy under the full-response energy distance and binary distributional accuracy are central goals, while also demonstrating the value of reporting complementary energy distances in mixed-response problems.

\bibliographystyle{apalike}
\bibliography{references}

\end{document}